\documentclass[letterpaper,10pt]{article}
\usepackage {osameet3}
\usepackage {amsmath,amssymb}
\usepackage [colorlinks=true,bookmarks=false,citecolor=blue,urlcolor=blue]{hyperref} 
\usepackage	{graphicx}
\usepackage {amsfonts}
\usepackage {algorithmic}
\usepackage	{threeparttable}
\usepackage	{multirow}
\usepackage {float}
\usepackage [caption=false]{subfig}
\usepackage {geometry}

\begin{document}
\title{Superscalar Parallel Carrier Phase Recovery with Transmitter
  I/Q Imbalance Compensation}
\vspace{-.2cm}
\author{Daniel R. Garc\'ia\textsuperscript{1} and Mario
  R. Hueda\textsuperscript{2,3}}

\address{
  \textsuperscript{1}Fundaci\'on Fulgor, Ernesto Ramagosa 518, C\'ordoba (5000), Argentina.\\
  \textsuperscript{2}Inphi Corp, C\'ordoba,
  Argentina. \textsuperscript{3}Digital Communications Research
  Laboratory, FCEFyN, UNC, Argentina.  }
\email{dgarcia@fundacionfulgor.org.ar; mario.hueda@unc.edu.ar}
\copyrightyear{2021}
\begin{abstract}\small
  We propose a superscalar parallel two-stage carrier phase recovery
  architecture to improve the performance of optical coherent
  receivers in the presence of Tx I/Q imbalance, Tx I/Q skew, and
  laser frequency fluctuations.
\end{abstract}

\vspace{.30cm}
\section{Introduction}
Decision-directed (DD) carrier phase recovery (CPR)
\cite{gianni_compensation_2013} and the blind phase search (BPS)
algorithm \cite{pfau_hardware-efficient_2009} have been widely used in
optical coherent receivers to compensate the impact of laser phase
noise. The two-stage CPR scheme based on a DD low latency parallel
phase locked loop (DD-PLL) followed by a feedforward CPR (e.g., BPS)
has demonstrated an excellent performance in the presence of laser
phase noise and frequency fluctuations. The latter are generated by
mechanical vibrations or power supply noise and modeled as a carrier
frequency modulation with a sinusoid of large amplitude (e.g., 200
MHz) and low frequency (e.g., 35 KHz)
\cite{gianni_compensation_2013}. CPR algorithms such as DD-PLL or BPS
use a symbol detector (or slicer) to estimate phase error. Therefore,
severe performance degradation may be experienced in the presence of
amplitude and phase imbalances introduced at the transmitter side,
between the in-phase (I) and quadrature (Q) components (i.e., Tx I/Q
\emph{imbalance}) \cite{zhang_algorithms_2019}. Tx I/Q time skew is
another impairment that degrades the receiver performance.

The compensation of the I/Q imbalance has been extensively addressed
in the literature (e.g.,
\cite{da_silva_widely_2016,lagha_blind_2020}). Typically, the
transmitter I/Q imbalance compensation at the receiver is achieved
\emph{after} the CPR stage. Unfortunately, the performance of this
approach with conventional CPR algorithms such as DD-PLL or BPS can be
seriously degraded with high-order modulation formats if severe Tx I/Q
imbalance is present. This degradation is a result of the
constellation warping at the input of the slicers in the CPR blocks
caused by the Tx I/Q imbalance \cite{zhang_algorithms_2019}. To deal
with this problem, some blind estimation schemes of the Tx I/Q
imbalance have been recently reported in the literature
\cite{zhang_algorithms_2019},\cite{zhang_modulation-format-transparent_2019}. Although
these proposals can achieve good compensation of the Tx I/Q imbalance,
their proper performance in the presence of laser frequency
fluctuations is challenged, owing to the large latency introduced by
their parallel implementation in high-speed transceivers
\cite{zhang_modulation-format-transparent_2019}, and the poor
performance of feedforward CPRs such as BPS
\cite{zhang_algorithms_2019} when carrier frequency fluctuations are
experienced \cite{gianni_compensation_2013}.

This work proposes a superscalar parallel (SSP) two-stage CPR to
improve the receiver performance in the presence of the Tx I/Q
imbalance, Tx I/Q skew, laser phase noise, \emph{and} carrier
frequency fluctuations. A first CPR stage based on a DD-PLL is used to
compensate frequency offset and carrier frequency fluctuations
\cite{gianni_compensation_2013}. The second stage, based on a
feedforward CPR (e.g., BPS), operates on the signal demodulated by the
DD-PLL and is mainly used to compensate the residual laser phase noise
not eliminated by the DD-PLL. The accuracy of the phase estimations
provided by both DD-PLL and BPS is improved by adding a one-tap
adaptive $2\times 2$ multiple-input multiple output (MIMO) equalizer
to compensate the Tx I/Q imbalance at the input of the slicers in the
CPR. As a result of the extra latency introduced by the one-tap MIMO
equalizer in the DD-PLL stage, the use of existing low latency
parallel DD-PLL schemes such as that proposed in
\cite{gianni_compensation_2013} is precluded for implementing in
high-speed receivers. Therefore, we use a superscalar parallel
architecture to reduce the latency of the DD-PLL
\cite{piyawanno_low_2010}. We show that the proposed SSP-based CPR
scheme with Tx I/Q imbalance compensation is able to drastically
improve the receiver performance even in the presence of Tx I/Q time
skew.

\section{Superscalar Parallel Two-Stage CPR with Tx I/Q
  Compensation}\label{sec:2}
\begin{figure}[!t]
	\centering
	\subfloat[]{\includegraphics[width=0.39\columnwidth]{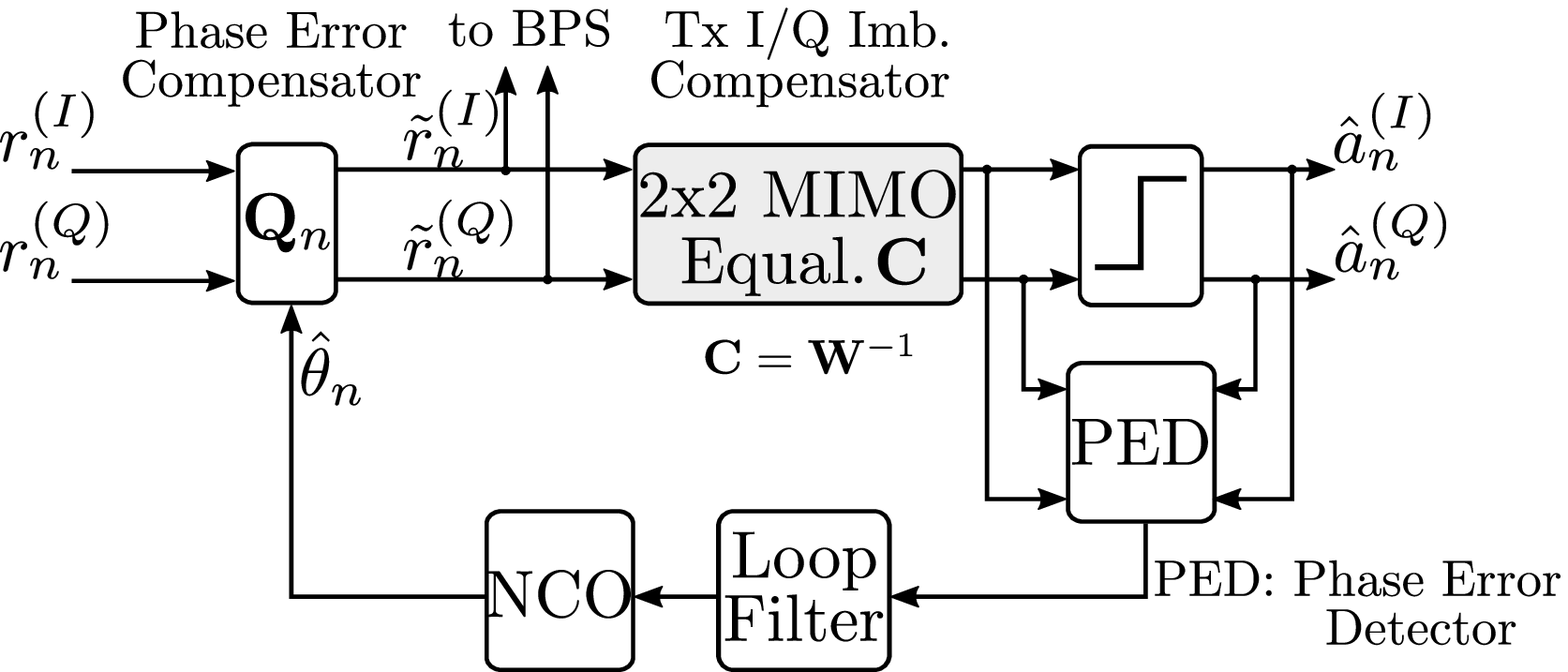}}
	\hfill
	\subfloat[]{\includegraphics[width=0.59\columnwidth]{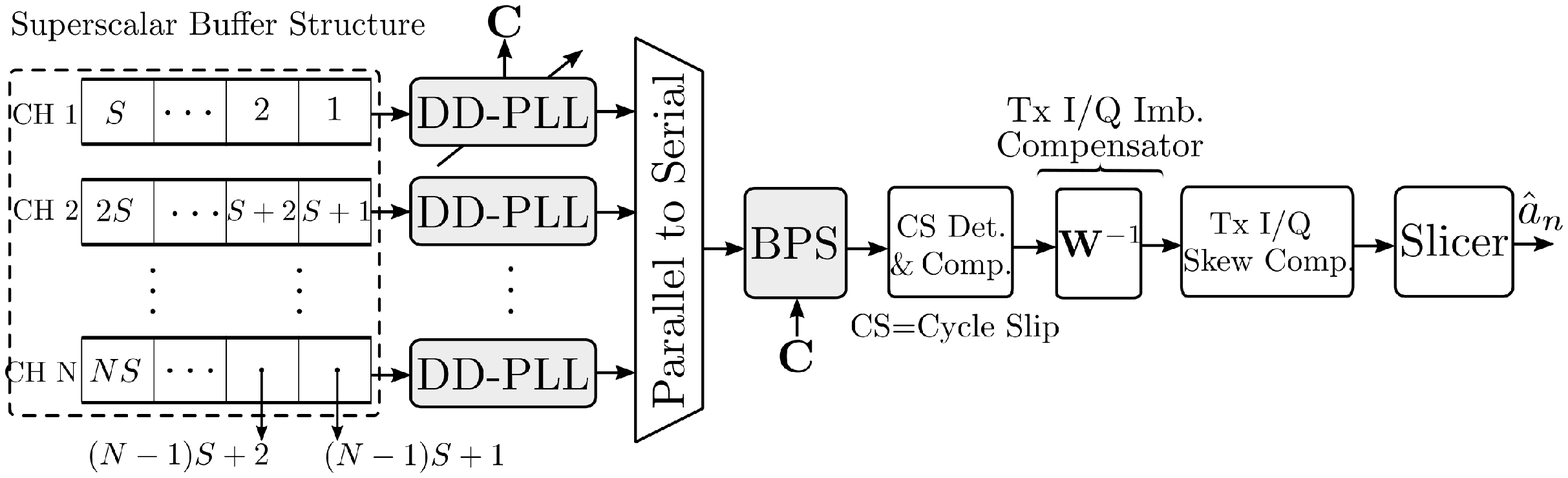}}
	\caption{\small Proposed (a) DD-PLL with Tx I/Q imbalance
       compensation and (b) 2-stage SSP-PLL+BPS.}
     \label{fig2_fig4a}
	\vspace{-.35cm}
\end{figure}
Without loss of generality, we consider the transmitted signal in one
polarization. The complex optical carrier with Tx I/Q imbalance can be
written as
$p(t)=
(1-\varepsilon_g)\cos[\omega_0t+\phi_e/2+\theta(t)]+j(1+\varepsilon_g)\sin[\omega_0t-\phi_e/2+\theta(t)],$
where $\varepsilon_g$ and $\phi_e$ are the gain and phase imbalance,
respectively, $\theta(t)$ is the carrier phase error, while
$\omega_0=2\pi f_0$ with $f_0$ being the carrier frequency
\cite{da_silva_widely_2016}. Let $a_n=a_n^{(I)}+ja_n^{(Q)}$ and $T$ be
the $n$-th transmitted complex quadrature amplitude modulation (QAM)
symbol and the symbol period, respectively. The equalized received
baseband signal (e.g., after chromatic dispersion compensation) in a
dispersive optical channel with Tx I/Q imbalance and carrier phase
error can be formulated as
\begin{equation}
\label{eq:2}
{\bf r}_n={\bf P}_n {\bf W} {\bf  a}_n+{\bf z}_n,
\end{equation}
where ${\bf r}_n=[r^{(I)}_{n}\quad r^{(Q)}_{n}]^{T_r}$ is a
$2\times 1$ real vector whose components are the received I/Q samples
(superscript $T_r$ denotes transpose),
${\bf a}_n=[a^{(I)}_{n}\quad a_n^{(Q)}]^{T_r}$,
${\bf z}_n=[z^{(I)}_{n}\quad z_n^{(Q)}]^{T_r}$ is the vector that
includes the amplified spontaneous emission (ASE) noise as well as any
other residual interference, while ${\bf W}$ and ${\bf P}_n$ are the
$2\times 2$ real matrices:
\begin{equation}
  \label{eq:W}
  {\bf W}=
  \begin{bmatrix}
    \cos(\frac{\phi_e}{2})  	&\sin(\frac{\phi_e}{2})	\\
    \sin(\frac{\phi_e}{2})  	&\cos(\frac{\phi_e}{2})	\\
  \end{bmatrix}
  \begin{bmatrix}
    1-\varepsilon_g  	&0 	  				\\
    0  	                &1+\varepsilon_g	\\
  \end{bmatrix};\quad
  {\bf P}_n=
  \begin{bmatrix}
    \cos(\theta_n)  	&-\sin(\theta_n)	\\
    \sin(\theta_n)  	&\cos(\theta_n)	\\
  \end{bmatrix},
\end{equation}
with $\theta_n=\theta(nT)$. Notice that $\bf W$ models the Tx I/Q
imbalance while ${\bf P}_n$ incorporates the effect of the phase
error, $\theta_n$. The latter is modeled as
$\theta_n=\psi_n+\Omega_cn+\Delta \Omega_n$, where $\psi_n$ is the
laser phase noise (i.e., $\psi_n=\sum_{k=-\infty}^n\eta_k$ where
$\eta_k$ are zero-mean iid Gaussian random variables with variance
$\sigma^2_{\eta}=2\pi T\Delta \nu$, being $\Delta \nu$ the laser
linewidth), $\Omega_c=2\pi T f_c$ with $f_c$ being the frequency
offset, while $\Delta \Omega_n$ is the phase change caused by the
frequency fluctuation
$\Delta \Omega_n=(A_p/\Delta f_c) \sin(2\pi T\Delta f_c n)$, where
$A_p$ and $\Delta f_c$ are the amplitude and frequency of the carrier
modulation tone \cite{gianni_compensation_2013}. Figure
\ref{fig2_fig4a}-a depicts the proposed first-stage CPR based on a
DD-PLL. The received signal \eqref{eq:2} is first demodulated by using
a $2\times2$ rotation matrix ${\bf Q}_n$ which uses the carrier phase
$\hat \theta_n$ provided by the DD-PLL (e.g., notice that
${\bf Q}_n={\bf P}^{-1}_n$ if $\hat \theta_n=\theta_n$). Then, the
samples are processed by a $2\times2$ real matrix $\bf C$ in order to
compensate the Tx I/Q imbalance at the slicer input (ideally,
$\bf C=\bf W^{-1}$). After that, the BPS algorithm estimates the phase
error by testing $B$ different phases. However, as pointed out in
\cite{zhang_algorithms_2019}, the Tx I/Q imbalance may cause BPS to
make wrong decisions and subsequently degrade the accuracy of the
phase estimation. To avoid this degradation, multiplication of the
slicer inputs by a $2\times2$ real matrix $\bf C$ is introduced in the
BPS to compensate the effects of the Tx I/Q imbalance, similarly to
what was done before with the DD-PLL as shown in
Fig. \ref{fig2_fig4a}-a.

\subsection{Impact of the Tx I/Q Time Skew}
The transmitter I/Q time skew degrades the receiver performance. This
effect is mainly caused by mismatches between the I/Q electrical path
responses from the Tx digital-to-analog converters (DACs) upto the
Mach Zehnder modulator (MZM). Based on the models used in
\cite{da_silva_widely_2016}, the discrete time electrical signal at
the MZM input in the presence of I/Q time skew can be written as
${\bf s}_n=\sum_{k}{\bf H}_k {\bf a}_{n-k}$, where ${\bf H}_k$ are
$2 \times 2$ real matrices whose elements depend on the I/Q skew,
$\tau$, and the Tx filter. Then, the received signal \eqref{eq:2} is
given by
${\bf r}_n={\bf P}_n {\bf W}_0 {\bf a}_n+{\bf z}_n+{\bf q}_n,$ where
${\bf W}_0={\bf W}{\bf H}_0$ while
${\bf q}_n={\bf P}_n {\bf W}\sum_{k,k\ne0}{\bf H}_k {\bf a}_{n-k}$ is
the $2\times 1$ real vector with the interference components caused by
the Tx I/Q time skew and the adjacent symbols. Therefore, the proposed
DD-PLL+BPS scheme with ${\bf C}={\bf W}_0^{-1}$ can compensate the Tx
I/Q imbalance (${\bf W}$) and a part of the Tx I/Q time skew
(${\bf H}_0$). Thus, the interference component ${\bf q}_n$ will be
\emph{seen} by the proposed DD-PLL+BPS as an extra noise component.

\subsection{Superscalar Parallel Implementation of DD-PLL (SSP-PLL)}
The high symbol rate requirements mandate the use of parallel
processing techniques for the implementation of coherent
transceivers. The tolerance of a conventional interleaving
parallelization system DD-PLL to the laser linewidth is reduced by a
factor $N\times D_L$, where $N$ is the parallelization factor and
$D_L$ is the processing delay (latency) due to the pipelined
implementation of the DD-PLL. A low latency parallel DD-PLL CPR has
been proposed in \cite{gianni_compensation_2013} to improve the
tolerance to the laser linewidth and frequency
fluctuations. Unfortunately, the benefits of this low latency PLL
\cite{gianni_compensation_2013} will be significantly reduced due to
the presence of the MIMO compensation filter ${\bf C}$ in the loop. To
deal with this problem, we propose to implement a superscalar
parallelization DD-PLL. The superscalar parallelization of the PLL has
been proposed in \cite{piyawanno_low_2010} for implementing a feedback
CPR in high-speed optical transceivers. SSP employs pilot
symbols\footnote{Pilot symbols of the SSP-PLL are used to correct
  cycle slips (CS) generated in the different CPR stages.}  and a
buffer to store the input samples, which are then rearranged to have
consecutive symbols in each parallelized channel (see
Fig. \ref{fig2_fig4a}-b). In this way, the processing delay of the
parallel architecture can be reduced from $N\times D_L$ to
$D_L$. Notice that a parallel implementation of all the DSP blocks
after the SSP-PLL stage (e.g., BPS, Tx I/Q imbalance and skew
compensation) is straightforward.

\section{Simulation Results}
\label{sec:3}
\begin{figure}[!t]
	\centering
	\subfloat[]{\includegraphics[width=0.66\columnwidth]{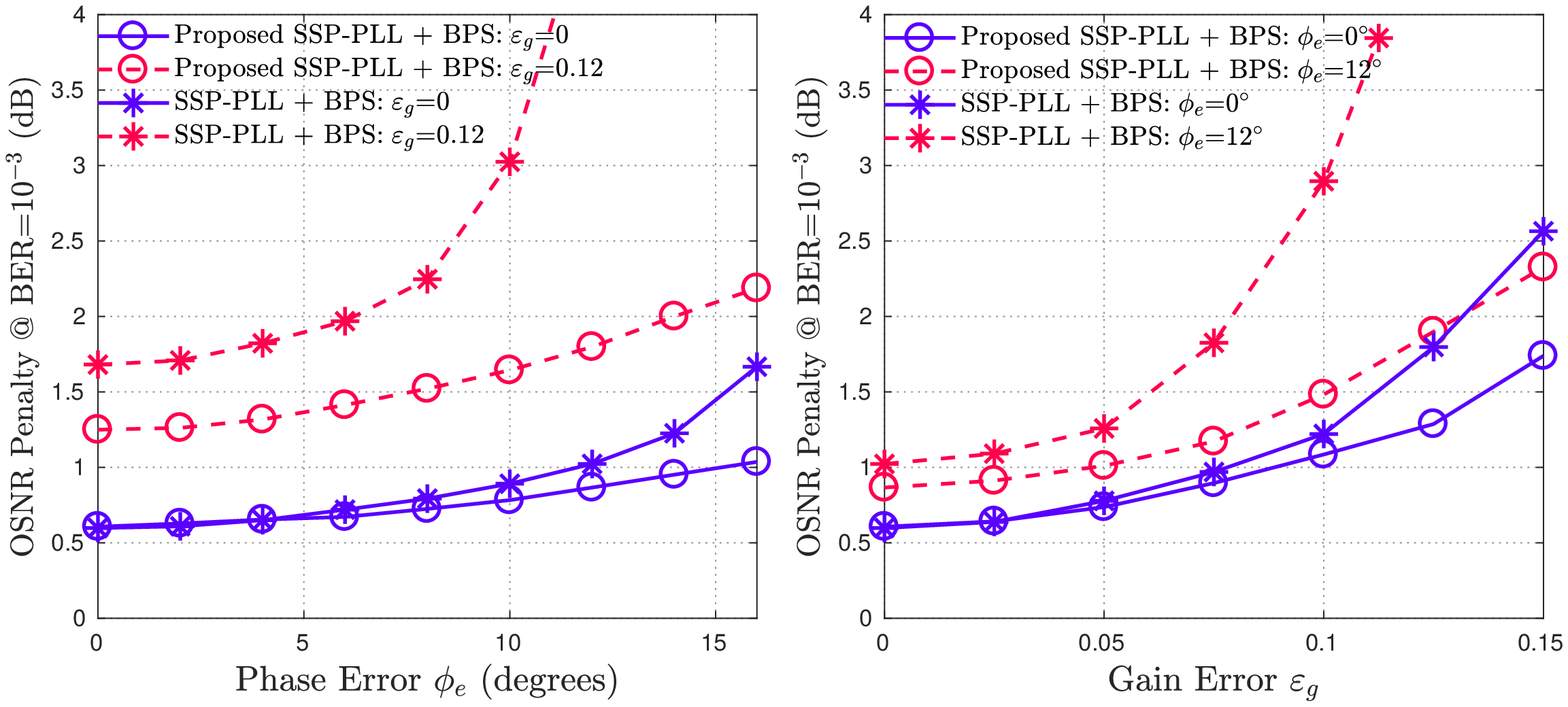}\label{fig5a}}
	\hfill
	\subfloat[]{\includegraphics[width=0.32\columnwidth]{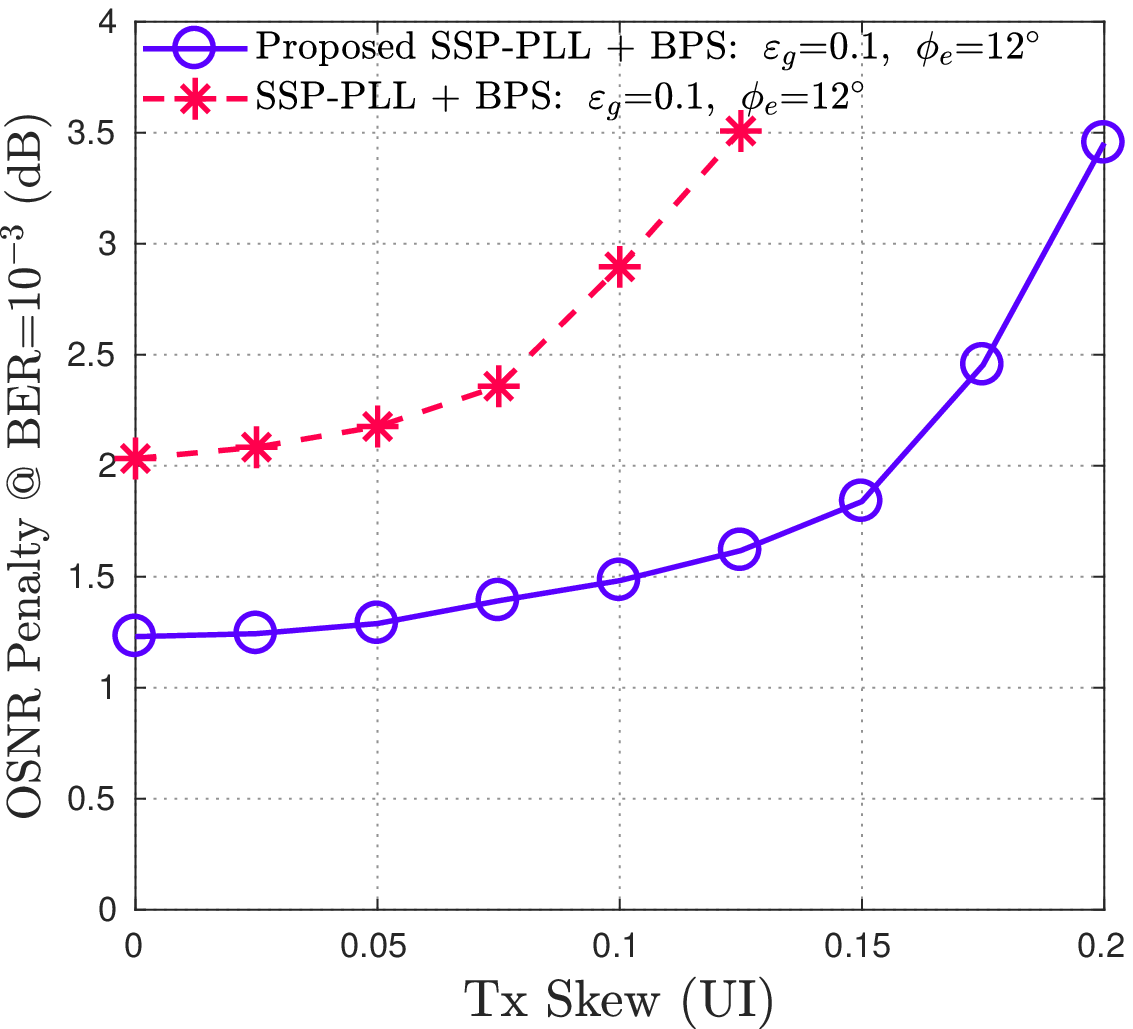}\label{fig6a}}
	\caption{\small (a) OSNR penalty versus $\phi_e$ and
       $\varepsilon_g$ with $\tau=0.1T$ for the proposed and
       conventional SSP-PLL+BPS. (b) Impact of the Tx I/Q time skew on
       the performance of the proposed and conventional SSP-PLL+BPS.}
	\label{fig5_fig6}	
     \vspace{-.65cm}
\end{figure}
We investigate the performance of the proposed two-stage CPR in the
presence of Tx I/Q imbalance and time skew, laser phase noise, and
carrier frequency fluctuations.  We consider 16-QAM with a baud rate
of $1/T=32$ Giga-baud (GBd), a type II second-order digital PLL with
$D_L=5$, and a BPS algorithm with filter length $M=40$ and $B=32$ test
phases.  The MIMO tap $\bf C$ is estimated at the receiver with the
LMS algorithm. The block buffer size is $N=16$ and $S=400$ with a
pilot overhead of 1\%. We consider $A_p=140$MHz, $\Delta f_c=35$KHz,
and $\Delta \nu=1$MHz. The Tx I/Q skew compensator at the Rx is
implemented by using two independent baud-rate adaptive real
equalizers.

Figure \ref{fig5_fig6}-a shows the optical signal-to-noise ratio
(OSNR) penalty at a bit-error-rate (BER) of $10^{-3}$ versus the Tx
phase and the gain imbalance with Tx I/Q time skew of
$\tau=0.1T$. Notice the important degradation caused by the Tx I/Q
imbalance in the conventional SSP-PLL+BPS solution.  In contrast, the
proposed two-stage CPR architecture is able to drastically reduce the
OSNR penalty at high values of gain and phase errors. Nevertheless, it
is interesting to highlight that the receiver performance worsens with
the increase of the Tx gain and phase imbalance even with the proposed
2-stage CPR and a perfect estimation of the MIMO compensation matrix
$\bf C$. This is caused in part by the ASE noise amplification
generated by the MIMO compensation filter since
$\det\{{\bf C}\}\approx\det\{{\bf W}^{-1}\}=1/\det\{{\bf W}\}>1$ if
$0<|\phi_e|<45^{\text o}$ and/or $0<|\varepsilon_g|<1$ (see
\eqref{eq:W}). Finally, Fig. \ref{fig5_fig6}-b depicts the OSNR
penalty at a BER of $10^{-3}$ versus the Tx I/Q time skew. Note that
the proposed SSP-PLL+BPS architecture for $|\tau|< 0.2T$ achieves an
important gain respect to the conventional CPR without Tx I/Q
imbalance compensation.

\vspace{-.15cm}

\section{Conclusions}
\label{sec:conclusions}
\vspace{-.10cm} A superscalar parallel CPR architecture with Tx I/Q
imbalance compensation for coherent optical receivers has been
proposed. Numerical results have demonstrated that the proposed
parallel CPR scheme is able to drastically improve the receiver
performance in the presence of Tx I/Q imbalance, Tx I/Q time skew, and
laser frequency fluctuations. This improvement will be more noticeable
when the modulation order is increased (e.g., 64-QAM).


\vspace{-.15cm}
\begin{thebibliography}{55}
  \small
\bibitem{morero_design_2016} D. Morero et al., ``Design Tradeoffs and
  Challenges in Practical Coherent Optical Transceiver
  Implementations,'' J. Light. Technol. \textbf{34}, 121--136 (2016).
\bibitem{gianni_compensation_2013} P. Gianni et al., ``Compensation of
  Laser Frequency Fluctuations and Phase Noise in 16-QAM Coherent
  Receivers,'' Photon. Technol. Lett. \textbf{2013}, 442--445 (2013).
\bibitem{pfau_hardware-efficient_2009} T. Pfau et al.,
  ``Hardware-Efficient Coherent Digital Receiver Concept With
  Feedforward Carrier Recovery for M-QAM Constellations,''
  J. Light. Technol. \textbf{27}, 989--999 (2009).
\bibitem{zhang_algorithms_2019} Q. Zhang et al., ``Algorithms for
  Blind Separation and Estimation of Transmitter and Receiver IQ
  Imbalances,'' J. Light. Technol. \textbf{37}, 2201--2208 (2019).
\bibitem{da_silva_widely_2016} E. da Silva et al., ``Widely Linear
  Equalization for IQ Imbalance and Skew Compensation in Optical
  Coherent Receivers,'' J. Light. Technol. \textbf{34}, 3577--3586
  (2016).
\bibitem{lagha_blind_2020} M. Lagha et al., ``Blind Joint Polarization
  Demultiplexing and IQ Imbalance Compensation for M-QAM Coherent
  Optical Communications,'' J. Light. Technol. \textbf{38}, 4213--4220
  (2020).
\bibitem{zhang_modulation-format-transparent_2019} Q. Zhang et al.,
  ``Modulation-format-transparent IQ imbalance estimation of
  dual-polarization optical transmitter based on maximum likelihood
  independent component analysis,'' Opt. Express \textbf{27}, 18055
  (2019).
\bibitem{piyawanno_low_2010} K. Piyawanno et al., ``Low complexity
  carrier recovery for coherent QAM using superscalar
  parallelization,'' \emph{in 36th Eur. Conf. and Exhib. on
    Opt. Commun.}, (Torino, Italy, 2010), pp. 1--3.
\bibitem{meng_jlt_2015} M. Qiu et al., ``Optimized Superscalar
  Parallelization-Based Carrier Phase Recovery for Agile Metro Optical
  Networks,'' J. Light. Technol. \textbf{34}, 1111--1119 (2015).
\end{thebibliography}
\end{document}